\documentstyle[12pt,epsfig,url]{article}

\begin{document}
~


\begin{center}
{\Large \bf Simplest Validation of the HIJING Monte Carlo Model}
\end{center}

\begin{center}
{\bf V.V.Uzhinsky\footnote{On leave from Laboratory of Information
Technologies of JINR, Dubna, Russia}\\ CERN, 2003}
\end{center}

\begin{center}\begin{minipage}{14cm}

Fulfillment of the
energy-momentum conservation law, as well as the
charge, baryon and
lepton number conservation is checked for the HIJING Monte Carlo
program in $pp$-interactions at $\sqrt{s}=$ 200, 5500, and 14000 GeV.
It is shown that the energy is conserved quite well. The transverse
momentum is not conserved, the deviation from zero is at the level
of 1--2 GeV/c,
and it is connected with the hard jet production. The deviation is
absent for  soft interactions. Charge, baryon and lepton numbers are
conserved.

~~~~~Azimuthal symmetry of the Monte Carlo events is studied, too. It is
shown that there is a small signature of a "flow". The situation with the
symmetry gets worse for nucleus-nucleus interactions.

\end{minipage}\end{center}

\section{Introduction}

A  Monte Carlo event generator server (GENSER) is created in the
framework of the LHC Computer Grid project (LHG) 
(see home page \url{http://lcg.web.cern.ch/LCG/}).
It is assumed that it
will collect well-tested generators adapted for the LHC energies and
experience of their application gained by the LHC collaborations. The
generators will be improved in the course of the LHC implementation,
test-beam experiments,
receiving new experimental data, and so on. The
server will give an opportunity for physicists to use the latest tested
generators versions.

The improvement of the generators means their adaptation to the
continuously changed computer technique, fine tuning of model
parameters, and inclusion of new physical effects such as collective
flow of particles, or black hole production. One of demands here is
the saving of results of previous studies, or determination of
their application
range. The change of the generators does not mean an
automatic change of the results of simulation of detector response, or
change of
the calibration constants of the experimental setups. The
generator changes are usually connected with inclusions of new physical
effects. For example, the problem of taking into account
collective flows of particles and jet quenching in the existing
nucleus-nucleus generators is very actual now after the RHIC
experiments. Clearly, the inclusion will not change the global
properties of simulated interactions dictated by soft and peripheral
collisions. At the same time, it can have  big influence on a small
probability of the hard jet production, or on the possibility of
registration of the collective flows. It is believed that the HIJING
event generator can be adapted most probably to these demands.

The HIJING model \cite{HIJING1} is the most popular one as applied for
simulation of nucleus-nucleus ($AA$) interactions at LHC energies.
It includes soft and hard interactions, nuclear modification of
structure functions, jet quenching, a true geometry of nuclear
collisions, and so on (description of the model see in
\cite{HIJING1,HIJING2}.
Thus, it is very important for future experiments to know its drawbacks
and accuracy of the model predictions. In the paper, the so-called
"theoretical" deficiency of the model will not be considered because it
is rather difficult to find a common point of view in the phenomenology.

The matter is that the authors of program-generators are solving a problem
of description of existing experimental data, on the one hand, and
on the other are
trying to introduce new theoretical ideas. As a rule,
the theoretical ideas are rather far away from everyday life, and it is
not easy to transfer them to observable predictions which are the most
important ones for the generators. In addition, all theoretical
approaches have areas of application, and unsolved questions. The
Monte Carlo event generator's authors have        problems
mainly with these
boundary conditions (let us point out only an unsolved problem of soft
and hard interaction connection). Thus, the Monte Carlo program
contains a lot of phenomenological things which can not be formulated,
in some cases, in a pure theoretical language.

We will concentrate on Monte Carlo program implementation of the model
presented in Ref.  \cite{HIJING2}, on its validation and installation.

\section{Code Installation}

The HIJING code located at
\url{ftp://nta0.lbl.gov/pub/xnwang/hijing} has been successfully
obtained by anonymous FTP. It contains 2 parts: {\it hijing1.383.f}
(the latest version) and {\it hipyset1.35.f}.
The first part collects the HIJING subroutines. The second part is
Pythia and Jetset7.2 subroutines adjusted for HIJING needs.
They must be compiled and linked together. {\it \bf hepf77} and {\it
\bf g77\_gcc2.32} compilers have been used.

It is assumed in the HIJING code that the
random number generator is called
{\bf\it RAN} with dummy argument {\it Nseed}. As  the function is
not automatically linked to the compilers, an additional function was
created and joined with the {\it main}
 program in order to run the HIJING
model at the AFS CERN system.
\begin{verbatim}
      FUNCTION RAN(NSEED)
      RAN=RLU(NSEED)
      RETURN
      END
\end{verbatim}
The function uses {\bf\it RLU} function from {\it hipyset1.35.f}.
The value {\it NSEED} was set to zero in introduced
{\it COMMON/RANSEED/NSEED}.

A run of the executable code was unpredicted at AFS CERN system.
Analysis has shown that local variables were not saved in some HIJING
routines. A typical example is presented in the function {\bf\it
romg}:
\begin{verbatim}
      FUNCTION ROMG(X)
C		********This gives the eikonal function from a table
C			calculated in the first call
      DIMENSION FR(0:1000)
      DATA I0/0/

      SAVE I0, FR                             !Uzhi

      IF(I0.NE.0) GO TO 100
      DO 50 I=1,1001
        XR=(I-1)*0.01
        FR(I-1)=OMG0(XR)
50    CONTINUE

100   I0=1
      IF(X.GE.10.0) THEN
        ROMG=0.0
        RETURN
      ENDIF
      IX=INT(X*100)
      ROMG=(FR(IX)*((IX+1)*0.01-X)+FR(IX+1)*(X-IX*0.01))/0.01
      RETURN
      END
\end{verbatim}

It was assumed that the array {\it FR} was filled, and the value of
{\it I0} was changed to "1" at the first call of {\it ROMG} function.
It was supposed that the array {\it FR} and the value of {\it I0} would
not be changed at all at following calls. This depends upon many
circumstances:  translator, operating system, computer type and so
on. At least at AFS system, the array {\it FR} was not saved. It is
sufficient to add a line marked by "{\it !Uzhi}" to protect a possible
error.

Such type  local variables have been found in the subroutines: {\it
jetini, romg, vegas, pystfe, pystfn}. "{\it SAVE}" operators have been
inserted in all of them The corrected code runs successfully at AFS
system and produces artificial events. The event record includes
multiplicity of particles, identificators of the particles, and their
kinematical characteristics ($E$, $P_x$, $P_y$, $P_z$). Only stable
particles were stored.

\section{Code Validation}

According to the existing imaginations, the events given by a Monte Carlo
event generator must satisfy the following demands:
\begin{enumerate}

\item Kinematical properties of the events must obey the
energy-momentum conservation law;

\item Conservation of charge, baryon and lepton numbers, strangeness
and so on must take place;

\item The events must have an azimuthal symmetry if unpolarized particles
are presented in an initial state;

\item The properties of the events and particles must be as close as
possible to known experimental data;

\item There must be a possibility to generate events at arbitrary
initial conditions in the range pointed by program authors, if there is
such restriction.

\end{enumerate}

Bellow we present the results of checking  the first three demands
for the HIJING program.

\subsection{$PP$-interactions}

Fig. 1 shows distributions on the sum of particle energies in the
events of $PP$-interactions at $\sqrt{s}=$200, 5500, and 14000 GeV in
the centre-of-mass system. As seen, in fact, the energy conservation takes
place.
\begin{figure}[cbth]
 \psfull
 \begin{center}
   \epsfig{file=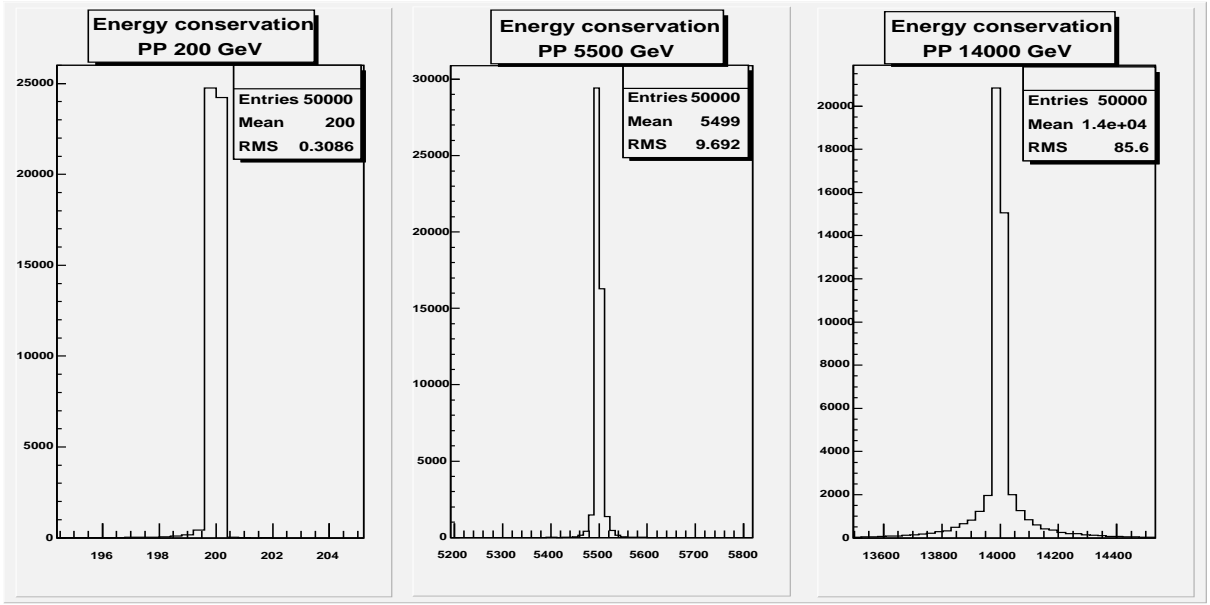,height=8.cm,width=16.cm}
 \end{center}
 \caption[fig1]{The energy distribution in the HIJING events}
 \label{fig1}
\end{figure}

Distributions on summarized momentum components of particles in the
events are presented in Fig. 2. The red histograms show the distributions
for $PP$-interactions at $\sqrt{s}=200$ GeV. The green ones give the
distributions at $\sqrt{s}=5500$ GeV (the energy of $NN$ interactions
in $AA$ collisions at LHC), and the blue histograms show the
characteristics at $\sqrt{s}=14000$ GeV. At first glance, there is not
a momentum conservation at all. Though the form of the distributions is
not a pure gaussian one. It seems that it consists of 2 distributions:
one is concentrated around zero, the other has large wings.

As it was said above, the HIJING program combined soft and hard
interactions. The soft interactions are simulated according to the
FRITIOF scheme \cite{FRITIOF} which saves energy and momentum. The hard
interactions are treated {\it a'la} Pythia algorithm. Thus, one can
expect different distributions for the processes. The calculations
presented in Fig. 3 confirm the expectation.

The black histograms in Fig. 3 show the distributions in the soft
interactions, and the yellow ones -- in hard collisions. As seen, the
transverse momentum is saved in the soft interactions.  The hard
interactions destroy the momentum conservation. There is no
longitudinal momentum conservation in both types of the interactions.
So, the problem of connection of soft and hard interactions is not
solved correctly in the HIJING program.

Of course, the problem of the momentum non-conservation can be solved
very easily for $PP$-interactions in the center-of mass system, as it 
is usually done in the Monte Carlo generators by the replacement of a 
particle momentum, $\vec p_i$, on $\vec p_i -1/N\sum_j \vec p_j$, where 
N is the multiplicity of the particles.  But the problem of the 
connection  remains.
\begin{figure}[t] 
 \psfull
 \begin{center}
   \epsfig{file=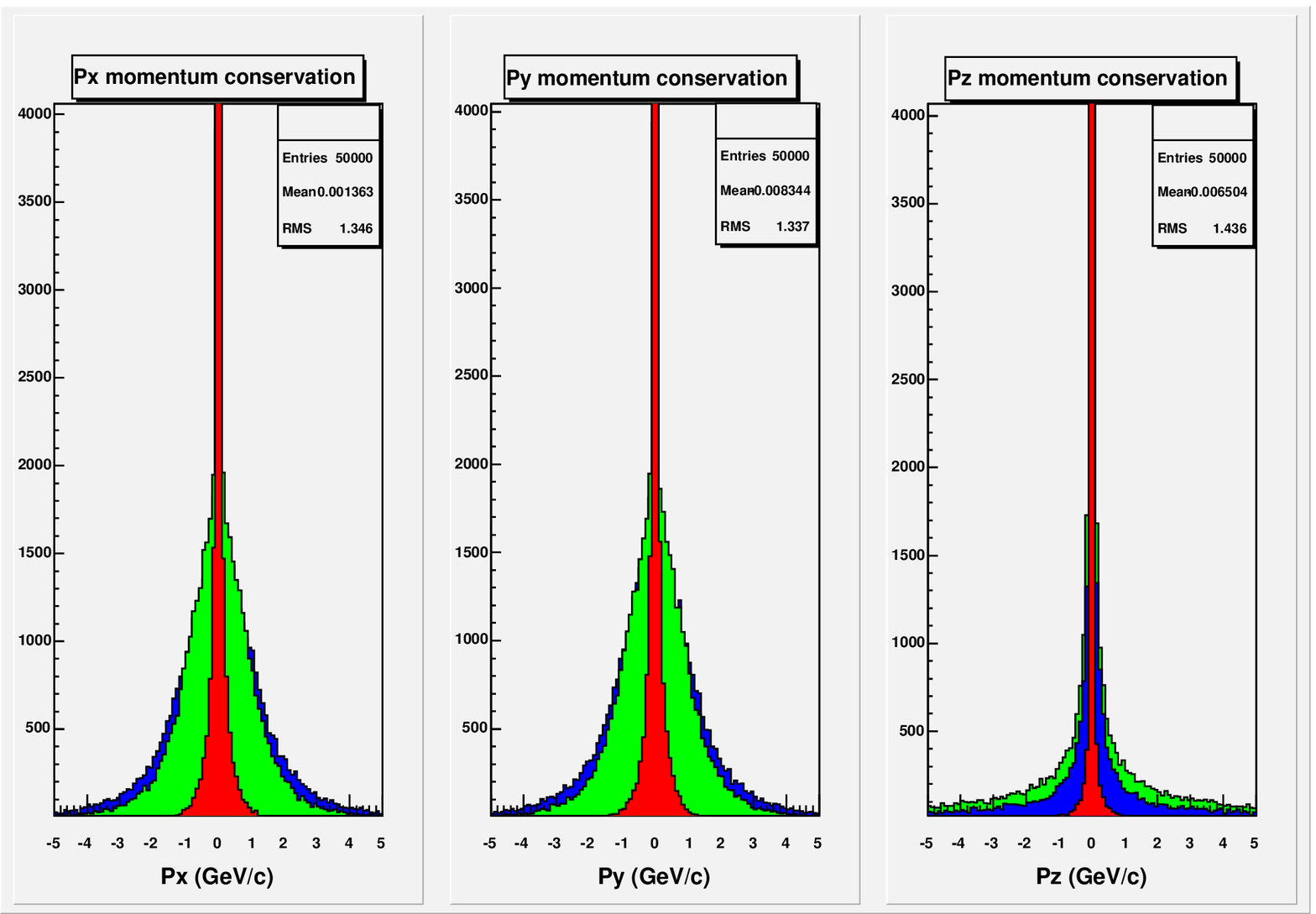,height=6cm,width=16.cm}
 \end{center}
 \caption[fig2]{The momentum conservation in the HIJING events}
 \label{fig2}
 \begin{center}
   \epsfig{file=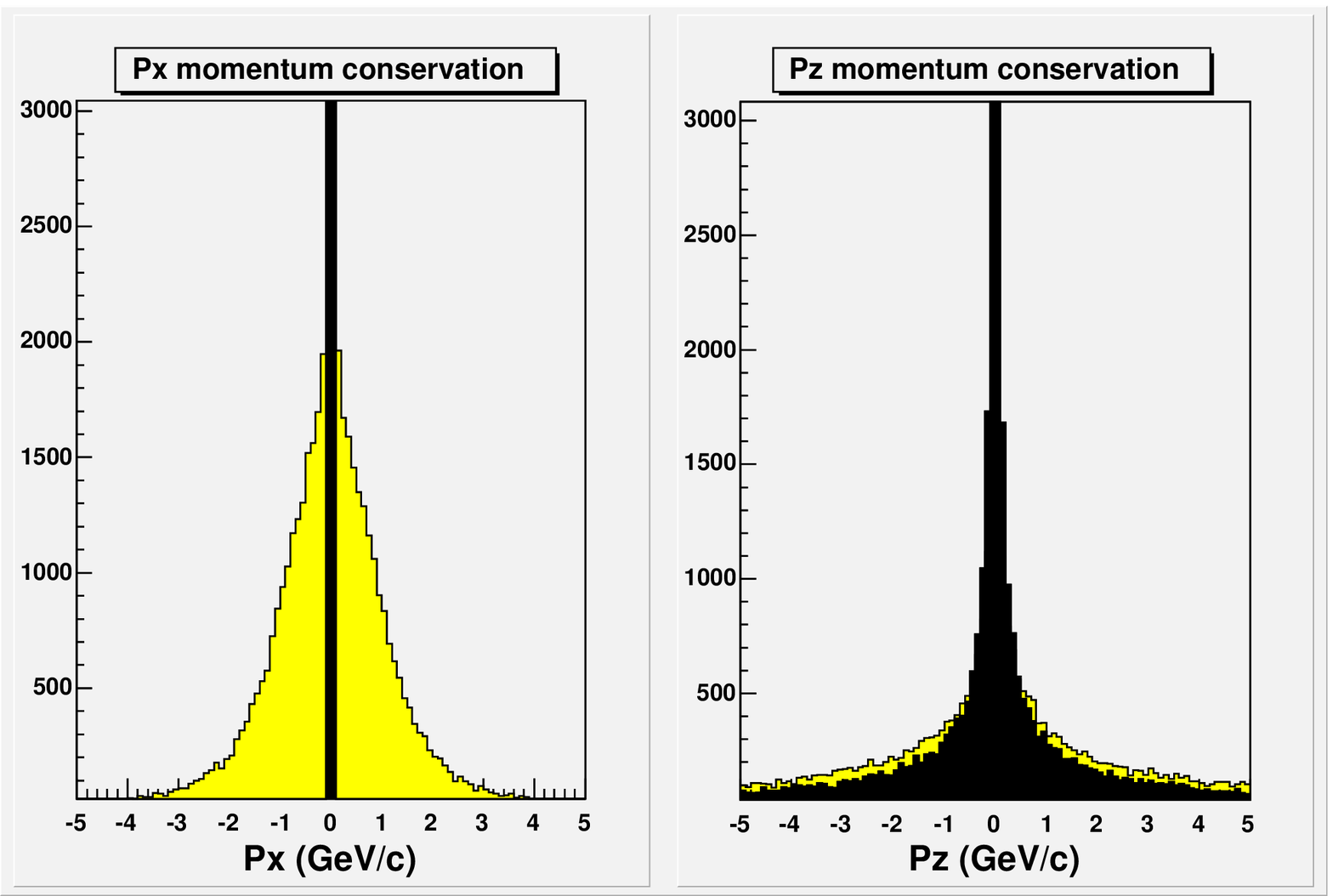,height=6cm,width=16.cm}
 \end{center}
 \caption[fig3]{The summarized momentum distributions at $\sqrt{s}=14000$
 GeV. The black histograms show the distributions in the soft
 interactions, the yellow ones -- in the hard interactions (with at
 least one pair of mini-jets).}
 \label{fig3}
\end{figure}

The non-conservation of the momentum can lead to an artificial collective
flow of the particles like a directed flow. In order to check the
possibility, azimuthal distributions of the particles have been
calculated. Fig. 4 presents small scale fluctuations of the
distributions.

The distribution at $\sqrt{s}=14000$ GeV shows quite a clear elliptic
flow pattern, though the value of $v_2$ is very small. At lower
energies the effect  disappears.  At the same time, the fluctuations
at all energies are  large enough
compared with pure statistical ones.
They can grow up in nucleus-nucleus interactions, and imitate
disoriented chiral condensate formation. The question must be
studied carefully in the future especially for the ALICE collaboration.

\newpage
\begin{figure}[cbth]
 \psfull
 \begin{center}
   \epsfig{file=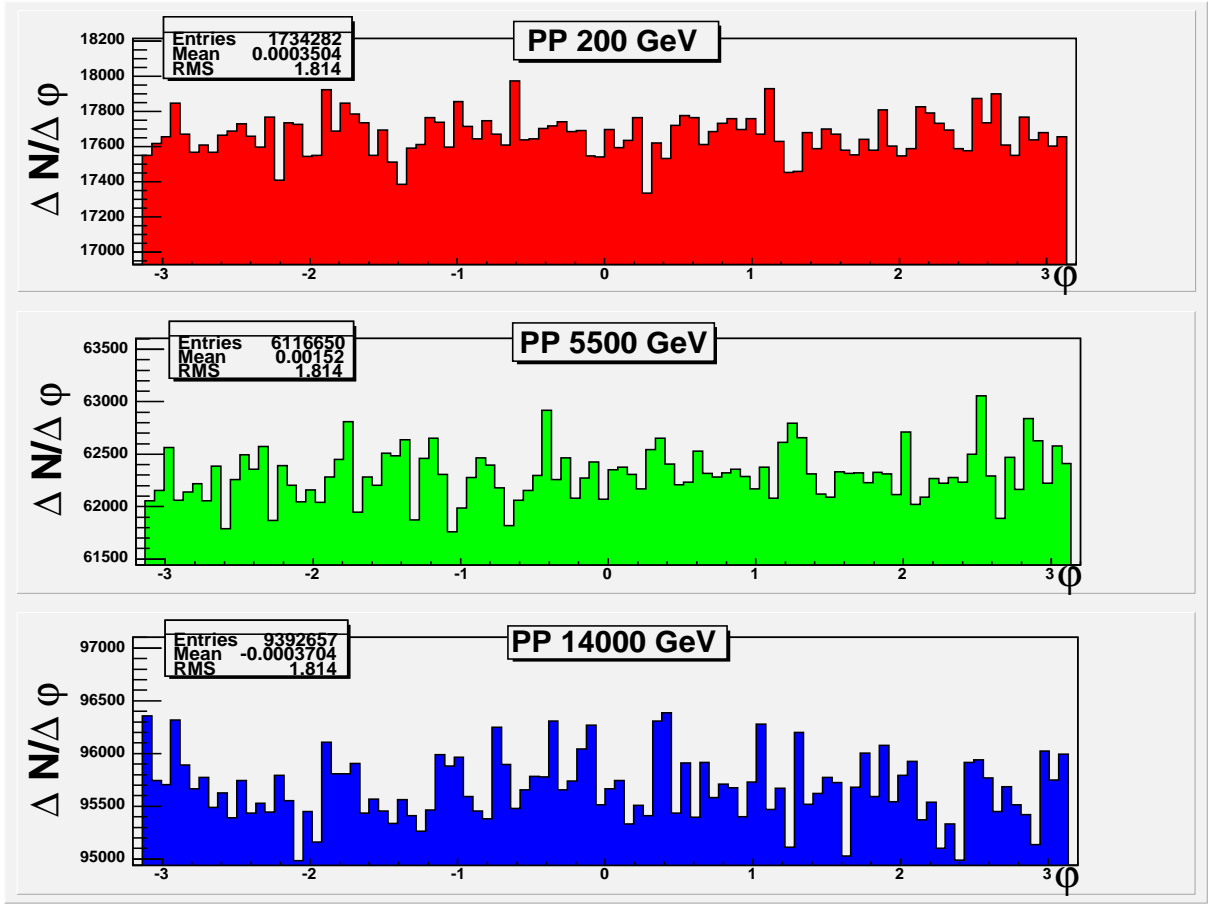,height=12.cm,width=16.cm}
 \end{center}
 \caption[fig4]{Azimuthal symmetry of the HIJING events}
 \label{fig4}
\end{figure}

\subsection{$AA$-interactions}

The HIJING program does not consider the usual nuclear effects such as
the Fermi motion of nucleons, relaxation of nuclear residuals,
absorption of mesons in nuclear matter, etc. Thus one can expect
that in nucleus-nucleus collisions the energy distribution in the
events will have  narrow peaks at $E=n*(E_{NN, cms}/2)$ where $n$ can
be 2, 3, 4, and so on. The first peak corresponds to the collision of one
nucleon from "projectile" nucleus and one nucleon from "target"
nucleus. The second peak  is connected with the collision of one nucleon
from the "projectile" nucleus and two nucleons from the "target" nucleus,
and vice versa.

The analogous peaks must be in the summarized longitudinal momentum
distribution at $Pz=\pm n*|P_{N, cms}|$ where $n$ runs from 0, 1, 2,
...

The calculation results presented in Fig. 5 for minimal bias
carbon-carbon interactions are in agreement with the expectations. If
the energy and the longitudinal momentum are conserved as in elementary
$NN$-collisions, the distortion of the peaks can not be seen in
the chosen scale.

A worse situation, as compared to that
in $PP$-interactions, takes place with
conservation of the transverse momentum. As seen, the yield of the
momentum saved component of the interactions becomes smaller. The
widths of the distributions increase. The widths of Px and
Py distributions are different, and this can be reflected on azimuthal
distribution of the particles.
\begin{figure}[cbth]
 \psfull
 \begin{center}
   \epsfig{file=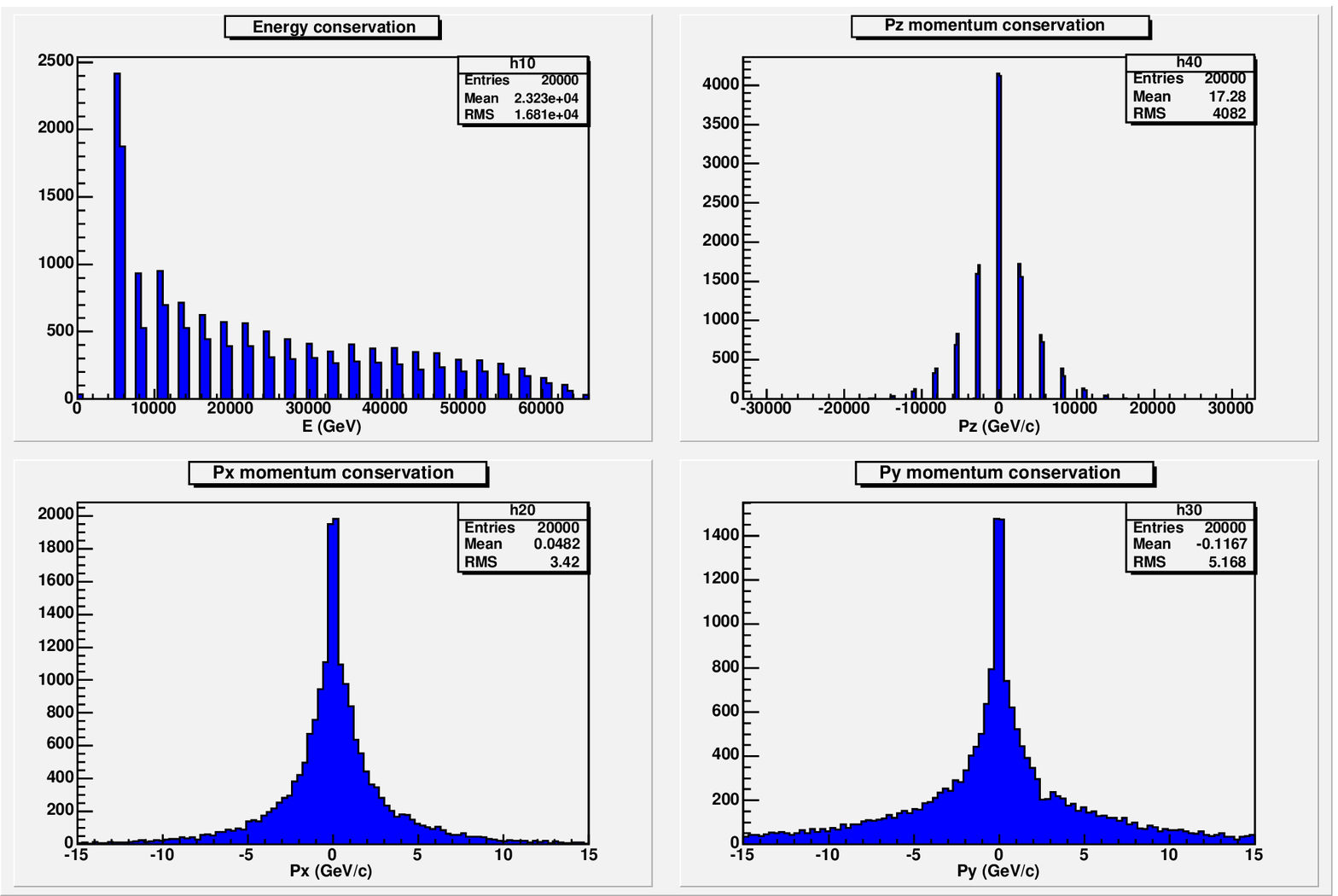,height=10.cm,width=16.cm}
 \end{center}
 \caption[fig5]{Properties of CC-interactions at $\sqrt{s_{NN}}=5500$ GeV
  in the HIJING model.}
 \label{fig5}
 \begin{center}
   \epsfig{file=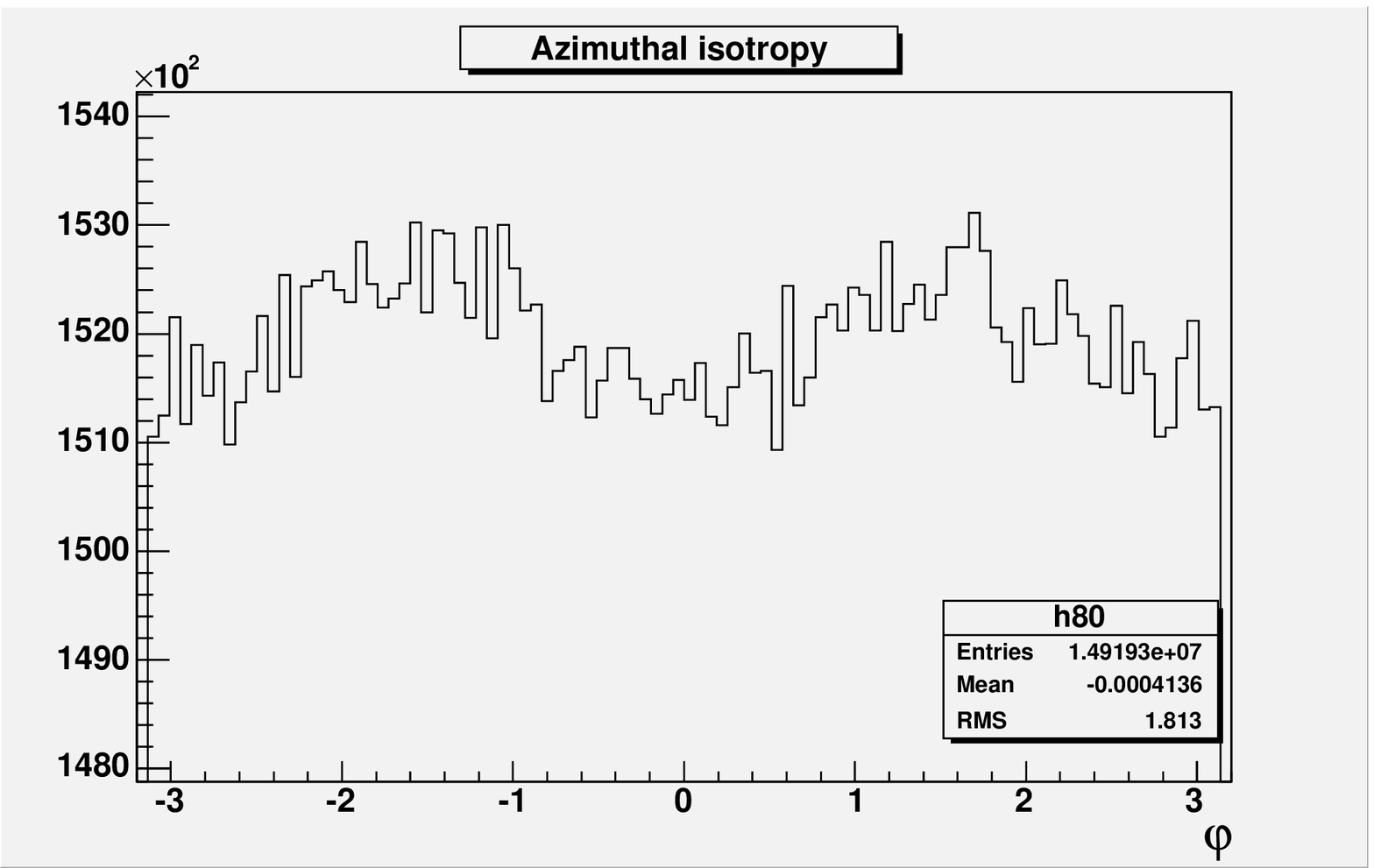,height=4.5cm,width=16.cm}
 \end{center}
 \caption[fig6]{Azimuthal distribution of particles in CC-interactions.}
 \label{fig6}
\end{figure}

Azimuthal distribution of the particles is shown in Fig. 6. As before,
only small scale fluctuations are shown. As seen, a clear "flow" signal
is observed! Maybe, the magnitude of the "flow" is too small, and this
drawback of the program is not important for the LHC experiment. At
least, the accuracy of the program must be checked up for $Pb+Pb$
interactions, where one can expect a larger "effect".

\section{Conclusion}
\begin{enumerate}

\item Energy is quite well conserved in the HIJING events.

\item Momentum is not conserved exactly in $PP$-collisions. The hard
interactions destroy the conservation. The violation of the momentum
conservation gets larger for the  AA-interactions.

\item The small "flow" pattern is presented in the azimuthal
distributions of the particles.

\end{enumerate}

The corrected HIJING program and codes used for this paper can be found
at\\
\url{http://lcgapp.cern.ch/cgi-bin/viewcvs/viewcvs.cgi/simu/GENSER/?cvsroot=Simulation
}

The author of the paper is thankful to A.~Morsch (ALICE) and P.~Nevski
(ATLAS) for useful discussions, P.~Bartalini, V.~Ivanov and P.~Zrelov
for support of the work, INTAS (grand N 00-0366) and RFBR (grand N 
03-02-17079) for their financial support.

\end{document}